\documentclass[prl,twocolumn,aps,amsmath,showkeys,showpacs,floatfix,nofootinbib,
reprint]
{ revtex4-1}

\usepackage{color}
\usepackage{graphicx}
\usepackage{bm}
\newcommand{\ignore}[1]{}
\def\beq{\begin{equation}}
\def\eeq{\end{equation}}
\def\beqa{\begin{eqnarray}}
\def\eeqa{\end{eqnarray}}

\newcommand{\braket}[2]{\left< #1 \big |  #2\right>}

\begin{document}

\title[Quantum Hall phases of two-component bosons]
{Quantum Hall phases of two-component bosons}

\author{Tobias Gra\ss$^1$, David Ravent\'os$^2$, Maciej Lewenstein$^{1,3}$, and Bruno Juli\'a-D\'iaz$^{1,2}$ }
\affiliation{$^1$ ICFO-Institut de Ci\`encies Fot\`oniques, 
Parc Mediterrani de la Tecnologia, 08860 Barcelona, Spain}
\affiliation{$^2$Departament d'Estructura i Constituents de la Mat\`{e}ria,
Universitat de Barcelona, 08028 Barcelona, Spain}
\affiliation{$^3$ ICREA - Instituci\'o Catalana de Recerca i Estudis Avan\c
cats, 08010 Barcelona, Spain}

\date{\today}
\begin{abstract}

The recent production of synthetic magnetic fields acting 
on electroneutral particles, like atoms or photons, has 
boosted the interest in the quantum Hall physics of bosons. Adding
pseudospin-1/2 to the bosons greatly enriches the scenario, as it allows them to
form an interacting integer quantum Hall (IQH) phase with no fermionic 
counterpart. Here we show that, for a small two-component 
Bose gas on a disk, the complete strongly correlated regime, 
extending from the integer phase at filling factor $\nu=2$ 
to the Halperin phase at filling factor $\nu=2/3$, is well 
described by composite fermionization of the bosons. 
Moreover we study the edge excitations of the IQH state, which, in agreement
with expectations from topological field theory, are found to consist of 
forward-moving charge excitations and backward-moving spin excitations. Finally,
we demonstrate how pair-correlation functions allow one to
experimentally distinguish the IQH state from competing states, like non-Abelian
spin singlet (NASS) states.
\end{abstract}

\maketitle

\textit{Introduction.} 
Recent progress in producing strong synthetic gauge fields in 
neutral systems like atomic quantum gases~\cite{dalibard,goldrev} 
or photonic fluids~\cite{carusotto} has catalyzed the research 
in bosonic quantum Hall states. While in the fractional quantum 
Hall (FQH) regime the bosonic states are often simply the counterparts 
of fermionic states, a significant difference occurs for 
non-interacting particles: Instead of forming an IQH liquid as fermions do, the
bosons' fate is to condense. 
However, as has been strikingly predicted by effective 
field theory~\cite{lu2012,senthil2013}, this does not exclude 
the possibility of an IQH phase of bosons. 
This phase can be obtained in a two-component system at 
filling factor $\nu=2$. As has been confirmed by numerical 
studies~\cite{furukawa2013,wu2013,regnault2013}, repulsive 
two-body contact interaction favors this phase against competing 
FQH phases. In contrast to the fermionic 
case, interactions are a crucial ingredient for the IQH physics of bosons.

Different from FQH states, IQH states have no anyonic
excitations, nor do they exhibit topological degeneracies in non-trivial
geometries (e.g. tori). Nevertheless, they possess topologically protected edge
states which due to 
Wen's edge-bulk correspondence~\cite{wen1992} make them distinct to
conventional bulk 
insulators. A particularly appealing property of the edge in 
spin-singlet systems is the fact that it can be excited in 
twofold ways: by spinless charge excitations (``holons'') 
or by charge-neutral spin excitations 
(``spinons'')~\cite{balatsky1991}. For the edge of the bosonic 
IQH phase, a $K$-matrix description predicts 
opposite velocities for these two types of excitations~\cite{senthil2013}, 
as a consequence of one positive and one negative eigenvalue of 
the $K$-matrix. This interesting property has been discussed before 
for a FQH state of spin-1/2 fermions at 
$\nu=2/3$ in a singlet~\cite{jmoore1997,wu2012}.

In the context of FQH physics, two-component 
Bose gases have been considered in a torus 
geometry~\cite{grass2012,furukawa2012}, where ground state degeneracies 
suggest them as a candidate for realizing NASS phases~\cite{ardonne-schoutens}.
Quantum many-body states 
with non-Abelian excitations are particularly relevant, as 
their use for topological quantum computations has been 
proposed~\cite{nayak}. A recent study of two-component bosons 
in a spherical geometry~\cite{wu2013}, however, gave rise to 
some controversy: Competitors of the NASS states are the composite 
fermion states which have Abelian topological order.
 
In this Letter we shed further light on the quantum 
Hall physics of two-component bosons by performing a systematic 
numerical study in a disk geometry. After briefly introducing different 
trial wave functions, we study for $N=6$ particles all incompressible 
states on the Yrast line, starting with the IQH 
state at $L_z=9$ (in units of $\hbar$) and ending with the Halperin 
state at $L_z=21$, where the system is able to fully avoid contact
interaction. We find all the incompressible states to be well
described 
by the composite fermion (CF) approach~\cite{jain-book}. We then study (for
$N=8$) the edge excitations of the IQH phase. Apart from some exceptions in the
forward-moving branch, we find number and spin of the edge excitations to
precisely agree with the predictions from effective theory.
A model of the edge states based on CF theory is 
found to accurately describe the wave functions of the backmoving states. It 
is shown that the forward moving states can be modeled by multiplying 
the ground state wave function with symmetric polynomials. Finally, we
demonstrate how pair-correlation functions distinguish the IQH state from
competing states in an experiment.

\textit{System and trial wave functions.} 
We study a system of $N$ two-component bosons described by the 
Hamiltonian 
$H=\sum_i^N \frac{[p_i-A(z_i)]^2}{2m} + 
\frac{m}{2} \omega^2 |z_i|^2 + V_0 \sum_{i<j} \delta(z_i-z_j)$, where 
$z_i=x_i+iy_i$ represents the position of the boson, 
$A(z)=\frac{B}{2}(x,-y)$ is a gauge potential, $m$, $V_0$, and 
$\omega$ are positive parameters specifying the mass, the two-body 
interaction strength, and the frequency of a 
harmonic confinement. The single-particle part of $H$ has a Landau level (LL)
structure, and is solved by Fock-Darwin (FD) states 
$\varphi_{n,\ell}$, which in the lowest Landau level (LLL) read 
$\varphi_{0,\ell}(z) \propto z^{\ell} \exp(-|z|^2/4)$, in units of length 
given by $\lambda=\sqrt{\hbar/(M\omega_0)}$, and 
$\omega_0 \equiv \sqrt{\omega^2 + \frac{B^2}{4m^2}}$.

One way to account for interactions is the CF 
theory developed by Jain~\cite{jain-book}. It provides a compelling 
picture to understand both IQH and FQH phases on 
an equal footing: By attaching magnetic fluxes to each particle, 
one obtains CFs, which are assumed to behave like 
non-interacting particles, that is, they fill one or several LLs. Originally,
this theory has been formulated for fermions, 
but it has been extended to bosonic quantum Hall phases triggered by 
the experimental progress in producing synthetic gauge fields 
acting on ultracold atoms~\cite{cooperwilkin}. Recently, 
CF states for two-component Bose systems have been 
introduced and studied in a spherical geometry~\cite{wu2013}.

Here we consider a two-component Bose system on a disk, for which 
CF states can be constructed in a similar way. Omitting the omnipresent 
Gaussian factor, we write~\cite{wu2013}, 
\beqa
\label{cf}
\Psi^{[n_a,n_b]}_{L_z} &=& {\cal P}_{\rm LLL} \left[ 
\Phi_{n_a}(\{z_a\}) \ 
\Phi_{n_b}(\{z_b\}) J(\{z\})\right]\,.
\eeqa
The last term is a Jastrow factor $J(\{z\})=\prod_{i<j}(z_i-z_j)$, 
which attaches one magnetic flux to each particle, turning the 
bosons into CFs. The wave function of the 
composite particles is given by the Slater determinants $\Phi_{n_a}$ and
$\Phi_{n_b}$, for particles of type $a$ and $b$, respectively. The indices
$n_{a(b)}$ yield the number of LLs occupied by the CFs. If $n_a=n_b$, the 
total spin is zero, $S=0$. Importantly, negative $n_a$ and $n_b$
shall refer to flux-reversed LLs: $\Phi_{-n}\equiv \Phi_{n}^*$.
Finally, ${\cal P}_{\rm LLL}$  projects back into the LLL of the bosonic system.
We perform this projection in the standard way by replacing
conjugate variables $z^*$ by derivatives $\partial_z$.

The only difference between Eq.~(\ref{cf}) and the corresponding 
definition on a sphere is the fact that in closed geometries 
the number of states in each LL is finite. This gives 
rise to the notion of ``completely filled'' LLs, and 
the state $\Psi^{[n_a,n_b]}$ is uniquely 
defined. Depending on the sign of $n=n_a+n_b$, its
filling factor is $\nu_{\pm}=n/(n\pm1)$. Contrarily, on a disk, there is more
than one way to distribute $N_a$ ($N_b$) particles in $n_a$ ($n_b$) LLs.
Typically each choice leads to a different total angular 
momentum $L_z$, such that wave functions at different angular 
momentum $L_z$ correspond to the same filling factor $\nu$ in 
the thermodynamic limit. Note that, for $|n_a|=|n_b|=1$, however, 
the wave functions are unique also on a disk. In particular, 
$\Psi^{[-1,-1]}$ has $L_z=N^2/4$ and corresponds to an integer 
filling factor, $\nu=2$. In contrast to all other CF wave functions 
with fractional filling, this wave function might describe an IQH liquid.

\begin{figure}[t]
\includegraphics[width=0.5\textwidth]{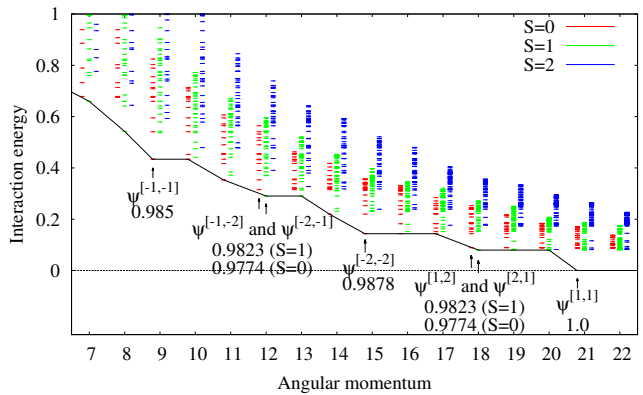}
\caption{(Color online)\label{yrast} Yrast line for $N=6$: 
For the incompressible states (marked by arrows) we give the overlap 
with corresponding CF wavefunctions.}
\end{figure} 

Another important trial wave function, obtained within  
the CF theory by putting all composite particles to the LLL ($\Psi^{[1,1]}$), is
the Halperin state \cite{halperin}, explicitly given by:
\begin{align}
\label{halperin}
&
\Psi_{\rm H} \sim
\prod_{i<j} (z_{ia}-z_{ja})^2
\prod_{i<j} (z_{ib-}z_{jb})^2
\prod_{i,j}
(z_{ia}-z_{jb}) \ .
\end{align}
It is a spin singlet wave function at filling $\nu=2/3$, with zero
energy in a two-body contact potential, and describes an Abelian FQH
phase. A series of non-Abelian quantum Hall states can be constructed from it
by forming $k$ clusters, putting each cluster into a Halperin state, 
and symmetrizing over all possible clusterizations~\cite{ardonne-schoutens}. 
In this way, one obtains the NASS states at filling factor 
$\nu=2k/3$ as the zero-energy eigenstates of $(k+1)$-body contact interaction.

\textit{Yrast line.} 
We have studied $N=6$ two-component bosons in the LLL on a disk by exactly
diagonalizing the SU(2)-symmetric 
two-body contact interaction. The presence of an additional 
harmonic trapping in $H$ which is invariant under spatial 
rotations along the $z$-axis and under spin rotations will not 
modify the eigenstates of the system, but simply increases the 
energy eigenvalues by a value proportional to $L_z$. Properly choosing 
the trapping frequency, one can tune the ground state of the system 
to different $L_z$.

The system's Yrast line, that is the spectrum of the interaction energy at fixed
$L_z$, is shown in Fig.~\ref{yrast}. Different $L_z=9,12,15,18,21$ correspond
to incompressible states, where an increase of angular momentum does not
directly lead to a decrease in energy. Notably, for all these $L_z$ it is 
possible to construct CF states. Moreover, exact ground states and CF states
agree in spin, and have very good overlap ($>0.97$). At $L_z=21$, the overlap
equals 1, as the Halperin state of Eq.~(\ref{halperin}) becomes the exact ground
state. At $L_z=18$, two CF states with $S_z\equiv (N_a-N_b)/2=0$ can be
constructed: $\Psi^{[1,2]}$ and $\Psi^{[2,1]}$. Accordingly, the ground state is
a triplet, but notably, also the antisymmetric combination of the two states
gives rise to a quasi-degenerate singlet state. For $L_z=15$, the CF
construction yields a unique singlet phase, $\Psi^{[-2,-2]}$, with overlap
0.9878 and large gap. For $L_z=12$, the situation is similar to $L_z=18$, with a
triplet ground state and a quasi-degenerate singlet state obtained from two
possible CF states, $\Psi^{[-1,-2]}$ and $\Psi^{[-2,-1]}$. The incompressible 
phase with smallest $L_z$ is found for $L_z=9$: the clearly gapped ground state
is a singlet and has large overlap (0.985) with $\Psi^{[-1,-1]}$. 

\textit{Edge physics of the IQH phase.} 
We now focus on this lowest-$L_z$ state on the Yrast line, for which 
we can extend our numerical study to $N=8$ particles and, accordingly, 
$L_z=16$. Compared to $N=6$, we find an only slightly smaller overlap,
$|\braket{\rm GS}{\Psi^{[-1,-1]}}|=0.9709$. As $\Psi^{[-1,-1]}$ 
describes a spin singlet with integer filling $\nu=2$, and the phase turns out
to be strongly gapped and incompressible, all prerequisites for an IQH phase
are fulfilled. Previous studies provided evidence of the integer topological
character of this phase by analyzing the properties of the entanglement entropy
on a sphere~\cite{furukawa2013,wu2013}, and the uniqueness of the ground 
state on a torus~\cite{regnault2013}. In the present paper, we consider 
the equivalent system in a plane, and focus on the physics at the edge to
characterize its topology~\cite{wen1992}.

An effective theory of the edge physics in fermionic singlet 
states~\cite{jmoore1997} is applicable also to the bosonic 
IQH state. It allows for a straightforward 
counting of the edge excitations. This theory is based on the observation that 
edge excitations of a spin singlet state might either be excitations 
which change angular momentum of the spin-up (down) particles, 
or be excitations which flip the spin of some particles. Thus, the 
effective edge Hamiltonian has the form~\cite{jmoore1997} 
$H_{\rm edge} \propto v_s (S_z^2 + \sum_l l b_l^{\dagger} b_l ) + v_c \sum_l l
c_l^{\dagger}c_l$. Here, the first term denotes the spinon excitations 
with velocity $v_s$, and the second term the holon excitations with 
velocity $v_c$. The operators $b_l$ and $c_l$ annihilate bosonic 
modes at angular momentum $l$.

An edge excitation at $|\Delta L_z| = 1$ can thus be achieved 
either by $\langle S_z^2 \rangle = 1$ and 
$\langle b_l^\dagger b_l \rangle = \langle c_l^\dagger c_l \rangle=0$, or 
by $\langle S_z^2 \rangle = 0$ and $\langle b_l^\dagger b_l \rangle =
\delta_{l1}$ 
and $\langle c_l^\dagger c_l \rangle=0$, giving rise to three states forming 
a spin triplet excitation, or by a spin singlet charge excitation with 
$\langle S_z^2 \rangle = 0$ and $\langle b_l^\dagger b_l \rangle= 0$ 
and $\langle c_l^\dagger c_l \rangle=\delta_{l1}$. In the case where $v_s<0$ 
and $v_c>0$, the spin triplet excitation corresponds to $\Delta L_z =-1$,
and the spin singlet to $\Delta L_z =+1$.

Extending this counting to excitations with $|\Delta L_z|>1$, we find 
that the multiplicities of the $c$ modes are given by the same counting 
which also applies to the Laughlin state, namely the number of positive-integer
sums which add up to $|\Delta L_z|$. From the effective theory, all these
lowest 
excitations are expected to be spin-singlets. Mixed charge-spin 
excitations would have higher energies. For $v_c>0$, these
modes are located at $\Delta L_z>0$. For the spin branch, the counting 
is less trivial: At $|\Delta L_z=2|$, four possible choices are possible, 
two with $\langle S_z^2\rangle=0$ and two with $\langle S_z^2\rangle=1$, 
thus giving rise to a triplet and a singlet. For $|\Delta L_z=3|$, two 
triplets and one singlet are expected, and for $|\Delta L_z=4|$, we expect 
two singlets, two triplets, and one SU(2) multiplet with total spin 
$S=2$. Again, mixed charge/spin excitations are expected at higher energies, and
for $v_s<0$ the spinon modes must have $\Delta L_z<0$.

\begin {figure}
\includegraphics[width=0.5\textwidth]{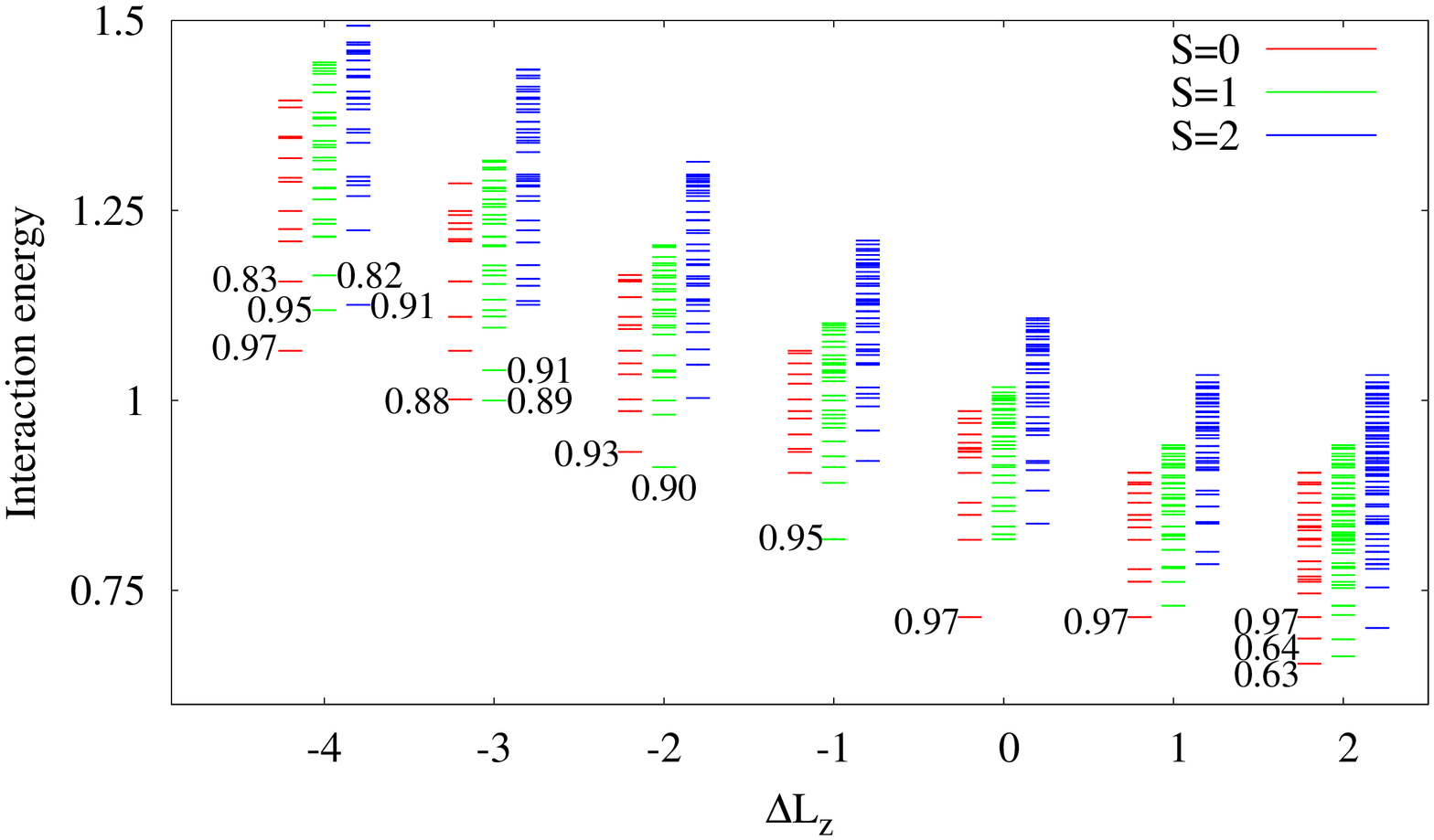}
\caption{(Color online)\label{edgespectrum} The spectrum of backward and
forward edge modes of a system with $N=8$ around $L_z=16$ ($\Delta L=0$). 
The numbers beside some states denote the overlap with trial wave functions
described in the text.}
\end {figure}

\textit{Backward moving edge states.}
In the spectrum shown in Fig.~\ref{edgespectrum}, we find one gapped 
triplet ground state at $\Delta L_z=-1$, and two quasi-degenerate gapped 
ground states, one singlet and one triplet, at $\Delta L_z=-2$. This 
perfectly matches with the counting expected from effective theory. 
Also at $\Delta L_z=-3$ and $\Delta L_z=-4$, the spin of the lowest states
agrees with the spin predicted by effective theory, but the degeneracy lifting
within the ground state manifold becomes larger than the gap to the excited
states. A particularly striking confirmation of the effective theory is the fact
that at $\Delta L_z=-4$ a $S=2$ multiplet becomes member of the ground 
state manifold.

A simple intuitive explanation for the presence of backmoving 
state, which directly leads to a scheme for constructing trial 
wave functions, can be given in terms of the CF 
approach: Since the ground state, $\Psi^{[-1,-1]}$, describes 
an IQH phase of CFs which are 
subjected to a flux-reversed magnetic field, a forward-directed edge
excitation of the CFs constitutes a backward-directed edge excitation of 
the bosons. More formally, as a consequence of the complex 
conjugation of the Slater determinants in $\Psi^{[-1,-1]}$, the edge
excitation of the CFs (that is the shift of one or several CFs 
to higher angular momentum) will correspond to a reduced angular
momentum of the bosons.

Following this reasoning, we have constructed trial wave functions 
for edge states with $-4 \leq \Delta L_z \leq -1$. For example, 
consider the state with $S_z=0$ at $\Delta L_z =-1$: The ground 
state $\Psi^{[-1,-1]}$ consists of four spin-up and four spin-down 
CFs, each filling the FD states with 
$\ell=0,\dots,3$ in the flux-reversed LLL. An 
edge state can then obtained in two ways: for either the 
spin-up or the spin-down CFs, we replace the 
FD state with $\ell=3$ by a FD state with $\ell=4$, 
which after complex conjugation leads to $\Delta L_z=-1$. Strikingly, 
after projecting these wave functions into the LLL, 
both choices lead to exactly the same wave function, and we recover a single
state at $S_z=0$, as demanded by both the effective theory and the numerical
results. 
This becomes more remarkable for $\Delta L_z <-1$: At $\Delta L_z =-2$, 
we find five ways to construct $S_z=0$ edge states, but they reduce 
to two linearly independent states. At $\Delta L_z =-3$, ten different 
constructions lead to three states, and at $\Delta L_z =-4$, twenty 
constructions yield precisely five different states. Thus, the 
CF construction is in perfect agreement with the 
counting of modes.
Apart from the counting, also the overlaps of the trial states 
with the exact states are remarkably high. They are explicitly 
given within Fig.~\ref{edgespectrum}, and for any of the eleven edge states 
in the interval $-4 \leq
\Delta L_z \leq -1$ they are larger than 0.82, demonstrating the
power of the CF description.

\textit{Forward moving edge states.}
For $\Delta L_z >0$, the effective theory predicts spin singlet ground states,
with degeneracy $1,2,3,5,\dots$ for $\Delta L_z=1,2,3,4,\dots$. Indeed we find
a single singlet ground state at $\Delta L_z=1$, though it is not separated by a
large gap from a second, low-lying triplet state, see Fig. \ref{edgespectrum}.
Also at $\Delta L_z=2$, there is a singlet ground state, but a nearby second
state in the spectrum is a triplet state, instead of a second spin singlet. At
$\Delta L_z=3$, even the ground state is a triplet. It has been argued that
forwardmoving edge states have a large velocity and thus merge with
bulk excitations, spoiling the spectral structure expected from effective theory
\cite{jmoore1997,wu2013}. Moreover, we note that the state $\Psi^{[-1,-1]}$ is
the first incompressible state on the Yrast line. Therefore, while backmoving
modes of this state do not interfere with forward moving edge modes of other
incompressible states, the forwardmoving excitations of $\Psi^{[-1,-1]}$
are expected to mix with backmoving modes of an incompressible triplet phase
at $L_z=20$ (for $N=8$).

Nevertheless, it is possible to identify some states in the spectrum of Fig.
\ref{edgespectrum} as forward moving edge states of $\Psi^{[-1,-1]}$. We
construct them by multiplying the ground state by homogeneous polynomials
which are symmetric in all variables. Such excitation might either act on the
bosons, that is on the wave function $\Psi^{[-1,-1]}$, or on the composite
fermions, that is on the CF wave function \textit{before} LLL projection.
Remarkably, the latter approach yields slightly better results.

For $\Delta L_z=1$, the construction yields one singlet, having overlap 0.9709
with the exact state. Note that this is precisely the overlap of the exact
ground state at $\Delta L_z =0$ with $\Psi^{[-1,-1]}$, suggesting that the
construction of the edge itself is exact, and the slight deviation of the
overlap from unity is caused by a discrepancy between the ground state at
$L_z=16$ and the CF state. Also, the ground states at both $\Delta L_z =0$ and
$\Delta L_z =1$ have exactly the same energy.

At $\Delta L_z=2$, the energy of only the sixth state in the spectrum, a
singlet, matches with the ground state energy at $\Delta L_z =0$. This state is
well reproduced (again overlap 0.9709) by our construction of edge states which
now yields two singlet states. At lower energies, we find two singlet
states, two triplet states, and one $S=2$ multiplet. Each of the two singlet
states has an overlap around 0.63 with our edge state construction, suggesting
that a linear combination of the two states would reasonably well agree. In that
way, we can, out of the three low-energy singlet states, recognize two as the
edge states predicted by effective theory.

\begin{figure}[t]
\includegraphics[width=0.28\textwidth, angle =-90, ]{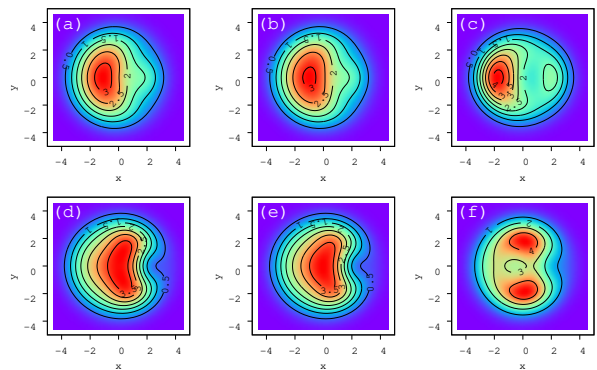}
\caption{(Color online)\label{pairco} Spin-dependent pair-correlation 
function for $N=8$ particles, showing the probability density 
of finding one particle with given spin after fixing another particle 
of the same spin (a--c) or of opposite spin (d--f) to $(x,y)=(2.5,0)$. In plots
(a) and (d), the system is in the exact ground state at $L_z=16$. in (b) and
(e), the system is in the corresponding CF state, and in (c) and (f), it
is in the $\nu=4/3$ NASS phase.}
\end{figure}

\textit{Experimental realization.}
The Hamiltonian studied here can be created in the 
laboratory by subjecting two-component bosonic atoms to artificial 
magnetic fields. Notably, such systems are flexible in size, and 
could be tuned from the microscopic regime (accessible by exact 
diagonalization) to the macroscopic regime (beyond exact diagonalization). This has
interesting applications: While a competition between the NASS state and the CF
state at filling factor $\nu=4/3$ takes place in the thermodynamic
limit~\cite{wu2013}, or for $N \geq 16$ on a disk, in smaller systems,
accessible to numerics, the two states occur at different $L_z$. The favored
phase could be determined, however, by an experiment. To illustrate this, let us
refer to a different competition which takes place for $N=8$ at $L_z=16$: We
have already seen that the CF picture with the $\nu=2$ state describes well the
ground state (overlap 0.97), but an alternative trial wave function is the
$\nu=4/3$ NASS 
state (overlap 0.52). Note that the CF state and the NASS state themselves 
have overlap 0.41, despite their different topological order. The overlaps
certainly give a clear picture in favor of the CF state, but they are not
accessible to experiment. What can be measured instead, are pair-correlation
functions, that is, the probability distribution of finding one particle 
somewhere in space, after another particle has been fixed 
at a given point. As shown in Fig.~\ref{pairco}, the pair-correlation functions
well distinguish between CF and NASS state, and a measurement of them would
be able to identify the ground state.

\textit{Conclusions.}
We have studied quantum Hall phases of two-component bosons on a 
disk. All incompressible phases are understood in the CF picture. The edge
states identify the IQH phase of bosons. This phase could be realized in
experiments with cold atoms, and detected by measuring pair-correlation
functions.

\textit{Acknowledgements}
This work has been supported by EU (SIQS, EQUAM), ERC (QUAGATUA), Spanish MINCIN
(FIS2008-00784 TOQATA), Generalitat de Catalunya (2009-SGR1289), and Alexander
von Humboldt Stiftung. BJD is supported by the Ram\'on y Cajal program.


\end{document}